\documentclass[aps,twocolumn,pra,superscriptaddress,amsmath,showpacs,tightenlines,pdflatex,longbibliography,10pt,conference]{revtex4-1}
\usepackage{amssymb}
\usepackage{amsmath}
\usepackage{dcolumn}
\usepackage{graphicx}
\usepackage{mathrsfs}
\usepackage{appendix}
\usepackage{graphicx}
\usepackage{booktabs}
\usepackage{color}

\newcommand{\ket}[1]{\mbox{$|#1\rangle$}}
\newcommand{\bra}[1]{\mbox{$\langle#1|$}}

\newcommand{\cor}[1]{{g_{\rm ss}^{(#1)}(0)}}

\usepackage{url}
\usepackage[colorlinks]{hyperref}
\hypersetup{%
    plainpages=true,
    breaklinks=true,
    hypertexnames=false,
    pageanchor=true,
    colorlinks=true,
    linkcolor={blue},
    citecolor={red},
    urlcolor={blue},
    anchorcolor={black}
}

\begin{document}

\title{Unconventional phonon blockade via atom-photon-phonon interaction in
hybrid optomechanical systems}

\author{Mei Wang}
\affiliation{School of physics, Huazhong University of Science and
Technology, Wuhan 430074, China} \affiliation{Theoretical Quantum
Physics Laboratory, RIKEN Cluster for Pioneering Research,
Wako-shi, Saitama 351-0198, Japan}

\author{Xin-You L\"{u}}
\email{xinyoulu@hust.edu.cn} \affiliation{School of physics,
Huazhong University of Science and Technology, Wuhan 430074,
China} 

\author{Adam Miranowicz}
\affiliation{ Faculty of Physics, Adam Mickiewicz University,
61-614 Pozna\'{n}, Poland}\affiliation{Theoretical Quantum Physics
Laboratory, RIKEN Cluster for Pioneering Research, Wako-shi,
Saitama 351-0198, Japan}

\author{Tai-Shuang Yin}
\affiliation{School of physics, Huazhong University of Science and
Technology, Wuhan 430074, China}

\author{Ying Wu}
\email{yingwu2@126.com} \affiliation{School of physics, Huazhong
University of Science and Technology, Wuhan 430074, China}

\author{Franco Nori}
\affiliation{Theoretical Quantum Physics Laboratory, RIKEN Cluster
for Pioneering Research, Wako-shi, Saitama 351-0198, Japan}
\affiliation{Department of Physics, The University of Michigan,
Ann Arbor, Michigan 48109-1040, USA}
\date{\today}

\begin{abstract}
Phonon nonlinearities play an important role in hybrid quantum
networks and on-chip quantum devices. We investigate the phonon
statistics of a mechanical oscillator in hybrid systems
composed of an atom and one or two standard optomechanical
cavities. An efficiently enhanced atom-phonon interaction can be
derived via a tripartite atom-photon-phonon interaction, where
the atom-photon coupling depends on the mechanical displacement
without practically changing a cavity frequency. This novel
mechanism of optomechanical interactions, as predicted recently by
Cotrufo \emph{et al.} [Phys. Rev. Lett. {\bf 118}, 133603 (2017)],
is fundamentally different from standard ones. In the enhanced
atom-phonon coupling, the strong phonon nonlinearity at a
single-excitation level is obtained in the originally
weak-coupling regime, which leads to the appearance of phonon
blockade. Moreover, the optimal parameter regimes are presented
both for the cases of one- and two- cavities. We compared
phonon-number correlation functions of different orders for
mechanical steady states generated in the one-cavity hybrid system,
revealing the occurrence of phonon-induced tunneling and different
types of phonon blockade. Our approach offers an alternative
method to generate and control a single phonon in the quantum
regime, and has potential applications in single-phonon quantum
technologies.
\end{abstract}

\pacs{42.50.Vk, 37.10.Vz, 37.90.+j} \maketitle

\section{Introduction}

Quantum effects in cavity optomechanical
systems~\cite{rOMS,OMS-Mart09,OMS-Asp12,OMS-Mey13,rCOMS08,Dong12}
and nanomechanical resonators
(NAMRs)~\cite{NEM03,Knobel03,NEM05,QM05} have attracted extensive
attention and progressed enormously in the past several years.
Reaching the quantum limits
of optomechanical systems is relevant for implementing
single-photon and single-phonon devices for quantum
metrology~\cite{QM92,QM99} and quantum information
processing~\cite{rQInformation1,rQInformation2,rQInformation3}. To
explore quantum effects of photons, many theoretical studies
(for a recent review see~\cite{Gu17}) have been focused on the
nonlinear quantum regime, where a single-photon nonlinearity is
comparable to a cavity decay. These studies were devoted, in
particular, to conventional~\cite{Tian92,Leonski94,Imamoglu98} and
unconventional~\cite{Leonski04,Miranowicz06,Liew10,Bamba11} photon
blockades, which are optical analogues of Coulomb blockade.

However, observing quantum effects in NAMRs becomes more complex.
The main obstacle is the thermal environment, which affects the
coherence and induces dissipation of NAMR states. In general, if a
NAMR is cooled to its ground state at low temperatures and the
frequency of the NAMR is high enough (of several GHZ), then the
oscillations of the NAMR quanta can beat the thermal energy and
approach the quantum limit. Even so, exploring purely quantum
phenomena for phonons is still very challenging, such as the
generation and detection of phonon
blockade~\cite{PBLiu16,PB16,Kimble05,Liao13,Rabl11,Shorcat09,Didier11},
the generation of nonclassical states of
phonons~\cite{Govia14,Usenko15}, including
 squeezed phonon
states~\cite{Hu96,Hu96-PRB,Hu97,Hu99} and Schr\"{o}dinger-cat
states~\cite{atom16}. Strong phonon nonlinearities become
necessary in order to explore the quantum behavior of NAMRs. To
satisfy this condition, NAMRs are often coupled  to an artificial
atom (or a qubit) to form a hybrid
system~\cite{Liu10,atom-resonator09,qubit-resontor10,rqubit,rhybrids,Pgate17,Singh10,atom-phonon09,atom-resonator08,Yi04,rSC1,Gu17}.
Such hybrid systems, with atom-phonon-interactions, have obvious
advantages for inducing strong phonon nonlinearities compared to
standard systems with phonon-phonon
interactions~\cite{Mahboob14,QED04} and photon-phonon
interactions~\cite{Lv15,Yin17,lvPra15}.
\begin{figure*}[t]
\centerline{\includegraphics[width=15.5cm]{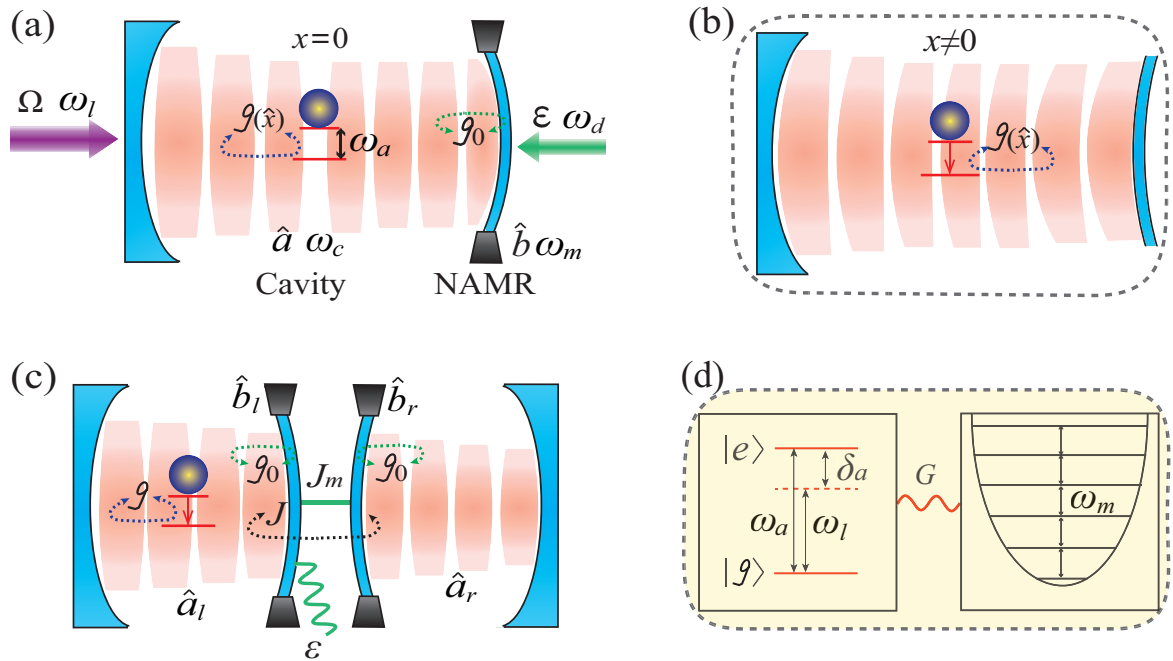}}
\caption{(Color online) (a) Schematic illustration of a
single-cavity hybrid optomechanical system, in which a two-level
atom is coupled to the cavity mode $\hat{a}$ of an optomechanical
system. The photon distribution in the cavity defines an electric
field, which is zero at the position of the atom when the
displacement $x=0$. To demonstrate quantum effects of the hybrid
optomechanical system, a strong driving field $\Omega$ with
frequency $\omega_l$ and a weak pumping field $\epsilon$ with
frequency $\omega_d$ act on the cavity mode (field $\hat{a}$ and
frequency $\omega_c$) and the NAMR (with field $\hat{b}$ and
frequency $\omega_m$), respectively. (b) Due to the displacement
of the NAMR at the cavity boundary, the electric-field
distribution at the atomic position becomes dense and induces an
atomic radiative transition. (c) A two-cavity hybrid
optomechanical system, where two standard optomechanical cavities
are coupled together, while a two-level atom is directly coupled
to the left cavity only. The two partially-transparent movable
mirrors (NAMRs) are in the middle, coupled to each other with
coupling rate $J_m$. The left NAMR is driven by a driving field
$\varepsilon$ with a resonant frequency $\omega_m$. (d) The
energy-level diagram of the reduced model for the hybrid
optomechanical systems in panels (a) and (c). In a rotating frame,
the atom couples to the NAMR with strength $G$, which is
controllable and plays a central role in these hybrid
optomechanical systems.} \label{fig:1}
\end{figure*}

To realize such atom-phonon interactions~\cite{Li16}, different
theories and experimental methods have been
presented~\cite{Connell10,atom-resonator08,Cleland04}. Strong
phonon nonlinearities can be obtained through a NAMR coupled
directly to an artificial atom, but it is difficult to be
engineered and dynamically controlled~\cite{Golter16,Wilson04}.
Recently, Cotrufo \textit{et al}.~\cite{Cotrufo17} predicted
a novel mechanism of optomechanical interactions, which
enables the indirect enhancement of the phonon nonlinearity via an
atom-photon-phonon (tripartite) interaction. This effect,
which is referred to as \emph{mode field coupling}, corresponds to
``a fundamentally different situation in which the mechanical
displacement induces a variation of the spatial distribution of
the cavity field, while cavity frequency is negligibly
affected''~\cite{Cotrufo17}. In other words, this mechanical
displacement modifies the cavity-field distribution which, in
turn, modulates the atom-phonon interaction. With this strong
controllable nonlinear interaction, many phenomena can be
explored, including phonon blockade and phonon-induced tunneling
in relation to their photonic analogues along the lines of,
e.g., Refs.~\cite{Miranowicz14,Liu14,Wang15}.

Phonon blockade (PB)~\cite{Liu10} refers to a quantum nonlinear
process, in which a single phonon of a nonlinear mechanical
resonator blockades the excitation of another phonon. This is a
close analogue of photon
blockade~\cite{Tian92,Leonski94,Imamoglu98}, which refers to a
process when a single photon in a nonlinear cavity blockades the
entry of another photon. Note that a Kerr-type mechanical
nonlinearity can simply be realized with a linear NAMR coupled to
a qubit in a Jaynes-Cummings-type model far-off resonance (i.e.,
in its dispersive limit)~\cite{Liu10}.

In systems which are more
complicated than the above-mentioned one, PB can occur as a result
of multipath interference. This effect is referred to as
unconventional PB (UPB), to emphasize this new mechanism of its
generation. UPB was predicted for, e.g., two coupled NAMRs with
two qubits~\cite{PBLiu16} and a coupled qubit-NAMR-cavity
system~\cite{Didier11}. This UPB is a direct analogue of
unconventional (or anomalous) photon blockade, which was first
predicted in Refs.~\cite{Leonski04,Miranowicz06,Liew10} and then
explained in terms of multipath interference in
Ref.~\cite{Bamba11} (for recent reviews see
Refs.~\cite{Gu17,Flayac2017}). Unconventional photon blockade can
occur for surprisingly \textit{weak} nonlinearities (even in the
\textit{weak}-coupling light-matter regime) and still enables very \textit{strong}
photon antibunching, as described by the second-order correlation
function $g^{(2)}(0)\approx 0$~\cite{Liew10}. Analogously, very
strong \emph{phonon} antibunching can be predicted for UPB, as
studied, in particular, here. We note that the terms of
unconventional photon and phonon blockades are \emph{not} only
used to refer to blockade in the weak-coupling regime (and, thus,
for weak nonlinearities). Indeed, these are also used for the
strong-coupling regime~\cite{Bamba11}.

The term UPB has also another meaning~\cite{Radulaski17},
which refers to unconventional properties of the generated field
via PB (as explained in Sec.~V), rather than to an unconventional
mechanism for creating PB or improving its quality.

Here we investigate UPB~\cite{PBLiu16} in a hybrid optomechanical
system featuring a controllable atom-phonon interaction obtained
from atom-photon-phonon interactions~\cite{Cotrufo17}. Physically,
this interaction results from a mechanically-induced modification
of the optical field distribution, which directly determines the
electric-field distribution around the atom. This leads to an
atom-photon displacement-dependent interaction generating
an atom-phonon interaction with an effective coupling rate $G$
($G'$). By applying a \textit{strong} external driving field to the cavity
and a \textit{weak} driving field to the NAMR, one can effectively control
the interaction between the atom and the NAMR. Thus, it is
possible to explore \textit{single-phonon} phenomena by utilizing such
atomic nonlinearity. With realistic parameters, we find
analytically and numerically an optimal regime for phonon
blockade.

Moreover, we also study atom-photon-phonon interaction resulting
from the optomechanical interaction in a two-cavity case. This
 tripartite interaction and atom-phonon interaction are
effectively enhanced in the two-cavity case in comparison to the
single-cavity case. We also calculate the steady-state and
time-variable phonon statistics. UPB is demonstrated with both
original and effective Hamiltonians leading to, effectively,
the same predictions. Thus, this paper offers an alternative
method to generate \textit{nonclassical phonon states through
modulated atom-phonon interactions}, and can have potential
applications in quantum technologies and quantum information
science.

\begin{figure}[t]
\centerline{\includegraphics[width=8.4cm]{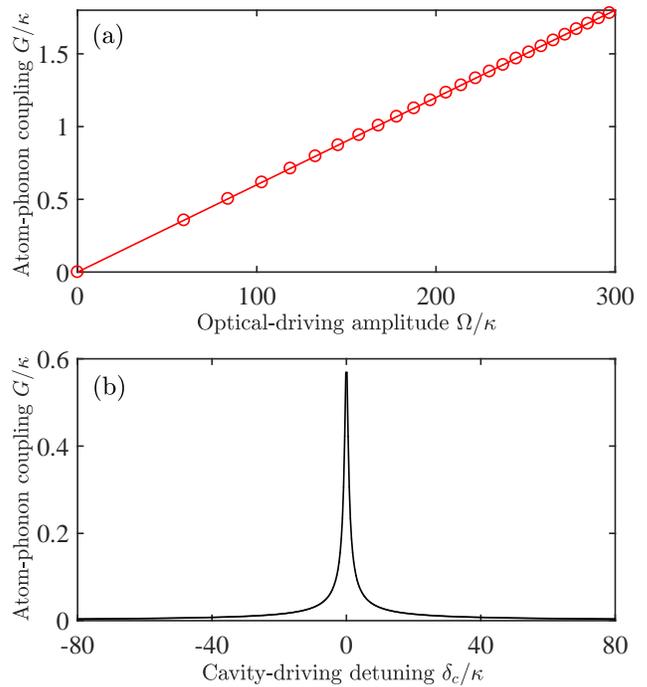}}
\caption{(Color online) The atom-phonon coupling strength
$G=\gamma\sqrt{n_{\rm cav}}$ versus: (a) the driving amplitude
$\Omega$ of the cavity mode with $\delta_c=0$ and (b) the detuning
$\delta_c$ between the cavity mode and the driving field with
$\Omega=100\kappa$ for the one-cavity hybrid system. The
parameters used here are $\omega_c/{2\pi}=470~\rm THz$,
$\kappa/{2\pi}=5~\rm MHz$, $\omega_m=280\kappa$,
$\Gamma_c=\kappa$, $\Gamma_{m}=0.01\kappa$, and
$\gamma=0.003\kappa$, all given in units of the atom-damping rate
$\kappa$.} \label{fig:2}
\end{figure}
This paper is organized as follows: In Sec.~II, we introduce a
one-cavity hybrid optomechanical system, in which an atom couples
to an optomechanical cavity mode. From the atom-photon-phonon
interaction in this system, we derive the atom-phonon interaction,
which is modulated by the average photon number in the cavity.
This illustrates that this atom-phonon interaction is controlled
by an external driving field. In Sec.~III, we discuss phonon
antibunching at low temperatures and give the corresponding
optimal conditions for observing phonon blockade in the one-cavity
system. In Sec.~IV, we implement the atom-photon-phonon
interaction from the optomechanical coupling in the system
consisting of two coupled standard optomechanical systems. In this
two-cavity system, the atom-phonon interaction is simultaneously
controlled by intra-cavity photon numbers and an
atom-photon-phonon interaction rate. Here we show the modulation
of the atom-photon-phonon interaction rate as a function of the optomechanical
coupling rate and the coupling rate of the two cavities. We also
demonstrate phonon blockade in this two-cavity system. Section~V
presents our analysis of higher-order phonon-number correlations
to reveal various types of phonon blockade and phonon-induced
tunneling in the single-cavity system. Our conclusions are given
in Sec.~VI.

\section{One-cavity system}

We now consider a hybrid optomechanical system, which is
schematically shown in Fig.~\ref{fig:1}(a). In this system, a
two-level atom is coupled to an optomechanical cavity, which, in
turn, interacts with a NAMR with coupling rate
$g_0$. The atom-photon coupling strength $g(\hat{x})$ is
displacement-dependent, which induces a variation of the spatial
distribution of the cavity field. Simultaneously, the photon
distribution around the atom determines the electric-field
intensity, which induces the radiative transition of the atom. As
shown in Fig.~\ref{fig:1}(b), when the mechanical displacement is
$x \neq 0$, then $g(\hat{x})\neq 0$ and the atom begins the
quantum Rabi oscillations between its ground and excited states.
Moreover, the cavity and mechanical modes are driven by a
monochromatic driving with frequency $\omega_l$ (amplitude
$\Omega$) and $\omega_d$ (amplitude $\epsilon$), respectively. In
a rotating frame with frequency $\omega_l$, the system Hamiltonian
under the rotating-wave approximation can be written
as~\cite{Cotrufo17}:
\begin{eqnarray}
\hat{H}_{1}/\hbar&=&\delta_c
\hat{a}^{\dagger}\hat{a}+\frac{\delta_a}{2}\hat{\sigma}_{z}+\omega_m
\hat{b}^{\dagger}\hat{b}+g(\hat{x})(\hat{\sigma}_{+}\hat{a}+\hat{\sigma}_{-}\hat{a}^{\dagger})
\nonumber\\&&
+\Omega(\hat{a}^{\dagger}+\hat{a})+\epsilon(\hat{b}^{\dagger}e^{-i\omega_d}+\hat{b}
e^{i\omega_d}), \label{e1}
\end{eqnarray}
where $\hat{\sigma}_{+}=\ket{e}\bra{g}$
($\hat{\sigma}_{-}=\ket{g}\bra{e}$) and
$\hat{\sigma}_{z}=\ket{e}\bra{e}-\ket{g}\bra{g}$ stand for the
raising (lowering) and occupation operators of the atom with Rabi
frequency $\omega_{a}$, respectively. Here $\ket{g}$ ($\ket{e}$)
is the ground (excited) state of the atom. Moreover, $\hat{a}$
($\hat{a}^{\dagger}$) and $\hat{b}$ ($\hat{b}^{\dagger}$) denote
the annihilation (creation) operators of the cavity mode (with
frequency $\omega_{c}$) and the mechanical mode (with frequency
$\omega_{m}$), respectively; $\hat{x}=\hat{b}^\dagger+\hat{b}$ is
the mechanical-mode position (displacement) operator, and the
detunings $\delta_{a,c}$ are defined by
\begin{eqnarray}
\delta_a=\omega_{a}-\omega_{l},
\quad\delta_c=\omega_{c}-\omega_{l}. \label{e2}
\end{eqnarray}
In the Hamiltonian~(\ref{e1}), we have neglected the
optomechanical interaction term, because the interaction rate is
 usually very weak. We assume that the frequencies satisfy the
condition $\omega_{a}=\omega_{c}+\omega_{m}$.

To better analyze the physical mechanism described by
Hamiltonian~(\ref{e1}), we now expand the photon-atom interaction
$g(\hat{x})$ to first order in the displacement,
\begin{eqnarray}
g(\hat{x})=g(0)+\gamma\hat{x},
\quad\gamma=\frac{dg(\hat{x})}{d\hat{x}}. \label{e3}
\end{eqnarray}
We assume that the effect of the displacement to the photon
distribution disappears at the point $x=0$, as depicted in
Fig.~\ref{fig:1}(a), leading to $g(0)=0$. Thus, substituting the
above expansion into Eq.~(\ref{e1}), we obtain the following
tripartite atom-photon-phonon interaction term
\begin{equation}
  \gamma\hat{x}(\hat{\sigma}_{+}\hat{a}+\hat{\sigma}_{-}\hat{a}^{\dagger}),
 \label{new1}
\end{equation}
where $\gamma$ is an interaction rate. This interaction is
different from the standard optomechanical coupling, because the
mechanical displacement changes the electric-field distribution in
the cavity, which induces atomic transitions without changing the
cavity frequency.

Under the conditions of strong optical driving, setting
$\delta_c=0$, we can obtain a large steady-state average photon
number $n_{\rm cav}$. The operator of the cavity field can be
written in the form $\hat{a}=\sqrt{n_{\rm cav}}+\delta \hat{a}$.
Combined with this reshaped operator and the tripartite
interaction, Hamiltonian~(\ref{e1}) can be transformed into a
driven Jaynes-Cummings-type Hamiltonian:
\begin{eqnarray}
\hat{H}'_{1}/\hbar&=&\frac{\Delta}{2}\hat{\sigma}_{z}+\Delta
\hat{b}^{\dagger}\hat{b}+G(\hat{\sigma}_{+}\hat{b}+\hat{\sigma}_{-}\hat{b}^{\dagger})
+\epsilon(\hat{b}^{\dagger}+\hat{b}), \label{e4}
\end{eqnarray}
in the rotating frame with frequency $\omega_d$, where
$\Delta=\delta_a-\omega_d$ is a detuning, and
$G=\gamma\sqrt{n_{\rm cav}}$ is an effective Jaynes-Cummings
interaction strength~\cite{PBLiu16,Zhang15, lvPra15}. Under the
condition $\gamma\ll G\ll\omega_m$, we have neglected the fast
oscillating terms with factors $\exp(\pm2i\omega_m t)$ in the
third term. Equation~(\ref{e4}) illustrates the physical mechanism
of the atom-phonon interaction shown in Fig.~\ref{fig:1}(d). The
harmonic oscillator with frequency $\omega_m$ is coupled to the
atom, which has a shifted frequency $\delta_a$ between the excited
and ground states in the rotating frame with frequency $\omega_l$.
Here the atom-phonon interaction $G$ is only controllable by the
photon occupation number $n_{\rm cav}$, because the
tripartite interaction rate $\gamma$ is a constant in the
one-cavity system. This enables us to control the atom-phonon
interaction by modulating the original optical-mode driving
strength. As one can see in Fig.~\ref{fig:2}, the coupling
strength $G$ is proportional to the amplitude of the driving
field, while the other parameters are constant. Furthermore, the
resonant driving of the cavity can drastically strengthen the
interaction strength $G$. These results illustrate that one
can easily control and enhance the atom-phonon interaction
$G$ from the weak to strong coupling regimes and even up to
the ultra-strong coupling regime. This also facilitates the
system to enter the quantum regime and display quantum properties
of phonons.

\section{Phonon blockade in the one-cavity system}

\begin{figure}[t]
\centerline{\includegraphics[width=8.6cm]{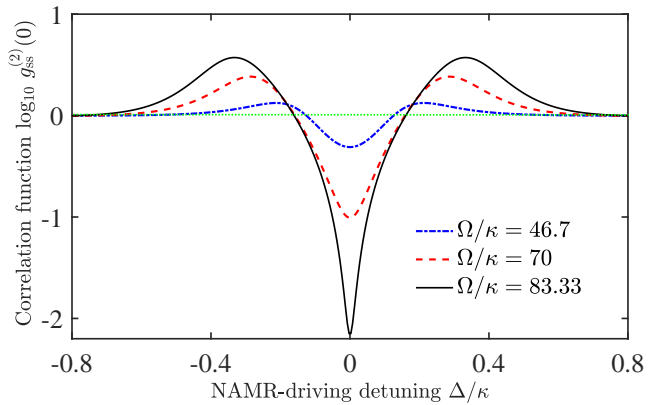}}
\caption{(Color online) Second-order phonon-number correlation
function $\cor{2}$ [precisely, $\log_{10}\cor{2}$] for the phonon
steady states generated in the NAMR versus the detuning $\Delta$
between the NAMR mode and the corresponding driving field with
different optical driving amplitudes $\Omega$ in the one-cavity
system. Here, the thermal phonon is $\bar{n}_{\rm
bth}=\bar{n}_{\rm ath}=0$ at low temperatures. The driving
amplitude of the oscillator is $\epsilon=0.01\kappa$, and the
detuning $\delta_c=0$, while all the other parameters are the same
as in Fig.~\ref{fig:2}.} \label{fig:3}
\end{figure}

Including the dissipation caused by the system-bath coupling, the
system dynamics is described by the Markovian master equation
\begin{eqnarray}
\dot{\hat{\rho}}&=&-i[\hat{H}'_{1},
\hat{\rho}]+\Gamma_m(\bar{n}_{\rm
bth}+1)\mathcal{D}[\hat{b}]\hat{\rho} \nonumber\\&&
+\Gamma_m \bar{n}_{\rm
bth}\mathcal{D}[\hat{b}^{\dagger}]\hat{\rho}+\kappa\mathcal{D}[\hat{\sigma}_{-}]\hat{\rho}, \label{e6}
\end{eqnarray}
where $\mathcal{D}[\hat{o}]\hat{\rho}=\hat{o}\hat{\rho}
\hat{o}^{\dagger}-(\hat{o}^{\dagger}\hat{o}\hat{\rho}+\hat{\rho
}\hat{o}^{\dagger}\hat{o})/2$ ($\hat{o}$ is a normal annihilation
operator) is the standard Lindblad dissipative superoperator for
the damping of the atom and NAMR. Here $\kappa$ and $\Gamma_m$ are
the damping rates of the atom and the NAMR, respectively.
Moreover, the thermal occupation of the atom has been ignored at very
low temperatures, i.e., $\bar{n}_{\rm ath}=0$. The thermal phonon
number $\bar{n}_{\rm bth}$ around the NAMR follows the
Bose-Einstein statistics $\bar{n}_{\rm bth}={\left[\rm
exp(\hbar\omega/\kappa_{B}\textit{T})-1\right]}^{-1}$, where
$\kappa_B$ is the Boltzmann constant and $\textit{T}$ is the
environment temperature.

\begin{figure}[t]
\centering
\includegraphics[width=8.4cm]{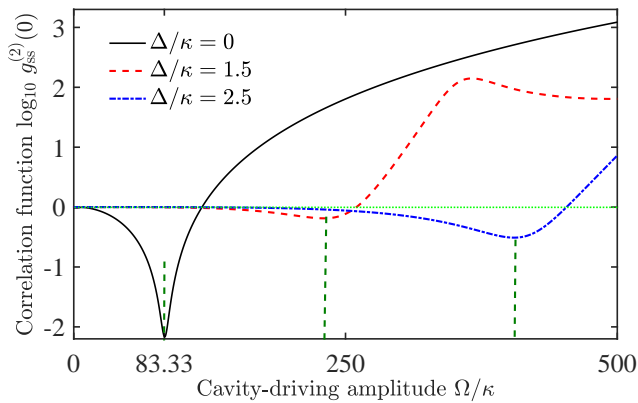}
\caption{(Color online) Correlation function $\log_{10}\cor{2}$
versus the optical driving amplitude $\Omega$ with different
detunings $\Delta$ between the mechanical mode and the
corresponding driving field in the one-cavity system. The
parameters used here are the same as in Fig.~\ref{fig:2}.}
\label{fig:4}
\end{figure}

By enhancing the atom-phonon interaction we find phonon blockade
in the one-cavity system. To observe quantum effects from a single
phonon, we numerically calculate the mechanical-mode
steady-state equal-time second-order correlation function
\begin{eqnarray}
g_{\rm ss}^{(2)}(0)=\frac{\langle
\hat{b}^{\dagger}\hat{b}^{\dagger}\hat{b}\hat{b}\rangle}{\langle
\hat{b}^{\dagger}\hat{b}{\rangle}^2}, \label{e7}
\end{eqnarray}
using the master equation in the basis of Fock states. The
numerical results shown in Fig.~\ref{fig:3} clearly
demonstrate that, assuming $\Delta=0$, the strongest phonon
blockade appears when $\Omega/\kappa=83.33$. At other points of
the optical driving amplitude, the value of the correlation
function becomes much larger even if the phonon blockade still
exists. The optimal point corresponds to
\begin{eqnarray}
G=\frac{1}{2}\sqrt{\kappa(\kappa+\gamma)}, \label{e8}
\end{eqnarray}
which is derived in Ref.~\cite{Xu16} from the wave function of the
truncated low-dimensional Hilbert space. Figure~\ref{fig:4} shows
the influence of the detuning $\Delta$ on the correlation function
$\cor{2}$, validating the optimal conditions specified
above.

Considering the energy of the system in the rotating frame, the
appearance of high-quality phonon blockade (as described by strong
phonon antibunching) is due to the resonance between a single
phonon and the shifted frequency of the atom. Then the atom-phonon
interaction induces the energy-level splitting of the system,
which blockades a second phonon entering into the system.

\section{Phonon blockade in the two-cavity system}

In this section, we analyze the atom-photon-phonon
interaction resulting from the optomechanical coupling in the
two-cavity system, shown in Fig.~\ref{fig:1}(c). Then we derive
the atom-phonon interaction, which is not only enhanced by the
average photon number, but also by the atom-photon-phonon
interaction rate $\gamma$. Finally, we predict phonon
blockade in the two-cavity system.

The schematic diagram of the two-cavity system is shown in
Fig.~\ref{fig:1}(c): two identical optomechanical cavities
(oscillators) $\hat{a}_l$ ($\hat{b}_l$) and $\hat{a}_r$
($\hat{b}_r$) are linearly coupled together with coupling rate $J$
($J_m$). The atom couples to the left cavity with coupling rate
$g$. The left mechanical mode is driven by a weak driving field
with resonant frequency $\omega_m$ and amplitude $\varepsilon$.
Then the Hamiltonian,
\begin{figure}[t]
\centering
\includegraphics[width=8.4cm]{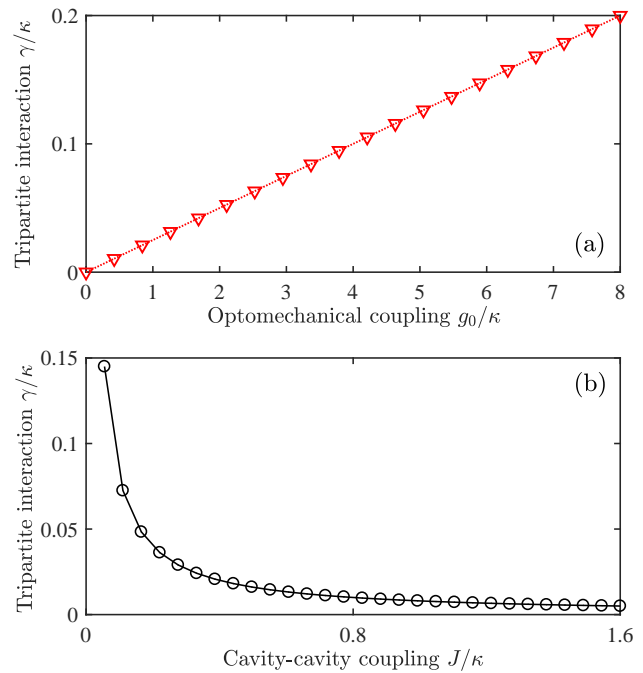}
\caption{(Color online) The atom-photon-phonon interaction rate
$\gamma$ in the two-cavity system versus: (a) the
optomechanical coupling strength $g_0$ with $J/{2\pi}=4~\rm MHz$,
(b) the coupling strength of the two cavities $J$ with
$g_0/{2\pi}=2~\rm MHz$. Here $g/{2\pi}=0.4~{\rm MHz},
J_m/{2\pi}=0.01~\rm MHz$, and all the other parameters are the
same as in Fig.~\ref{fig:2}.} \label{fig:5}
\end{figure}
representing the total two-cavity system reads as
\begin{eqnarray}
\hat{H}_{\rm
2tot}/\hbar&=&[\omega_c-g_0(\hat{b}_l^\dagger+\hat{b}_l)]\hat{a}_l^{\dagger}\hat{a}_l+
[\omega_c-g_0(\hat{b}_r^\dagger+\hat{b}_r)]\hat{a}_r^{\dagger}\hat{a}_r
\nonumber\\&&
+\frac{\omega_a}{2}\hat{\sigma}_z+\omega_m(\hat{b}_l^{\dagger}\hat{b}_l+\hat{b}_r^{\dagger}\hat{b}_r)+
J(\hat{a}_r^{\dagger}\hat{a}_l+\hat{a}_l^{\dagger}\hat{a}_r)
\nonumber\\&&
+J_m(\hat{b}_l^{\dagger}\hat{b}_r+\hat{b}_r^{\dagger}\hat{b}_l)+g(\hat{\sigma}_{+}\hat{a}_l+\hat{\sigma}_{-}\hat{a}_l^\dagger)
\nonumber\\&& +\varepsilon(\hat{b}_{l}^\dagger e^{-i\omega_m
t}+\hat{b}_{l}e^{i\omega_m t}).\label{e9}
\end{eqnarray}
To better analyze the physical mechanism described by this
Hamiltonian, we introduce the mechanical supermode annihilation
operators
\begin{equation}
  \hat{b}_{\pm}=\frac{1}{\sqrt{2}}(\hat{b}_l\pm \hat{b}_r).
 \label{supermodes}
\end{equation}
The supermode $\hat{b}_{+}$ interacts with the left and right
cavity modes with an equal coupling strength $-g_0/\sqrt{2}$,
while the supermode $\hat{b}_{-}$ interacts with these
cavities with an equal and opposite coupling strength $\mp
g_0/\sqrt{2}$. Then we can effectively omit the mode
$\hat{b}_{+}$ from Hamiltonian~(\ref{e9}). Moreover, the
condition $J_m\ll \omega_m$, enables us to safely neglect the
terms containing $J_m$. Under the condition of quasi-static
approximation~\cite{Cotrufo17} for $\hat{\delta}_b$, the optical
parts in Hamiltonian~(\ref{e9}) can be diagonalized in the
rotating frame with frequency $\omega_c$. Thus, the effective
Hamiltonian becomes
\begin{eqnarray}
\hat{H}_{2}/\hbar&=&\sqrt{\hat{\delta}_b^2+J^2}\hat{a}_{+}^{\dagger}\hat{a}_{+}-
\sqrt{\hat{\delta}_b^2+J^2}\hat{a}_{-}^{\dagger}\hat{a}_{-}+\frac{\omega_m}{2}\hat{\sigma}_z
\nonumber\\&&
+g\hat{\alpha}(\hat{\sigma}_{+}\hat{a}_{+}+\hat{\sigma}_{-}\hat{a}_{+}^\dagger)+g\hat{\beta}(\hat{\sigma}_{+}\hat{a}_{-}+\hat{\sigma}_{-}\hat{a}_{-}^\dagger)
\nonumber\\&&
+\omega_m\hat{b}_{-}^{\dagger}\hat{b}_{-}+\frac{1}{\sqrt{2}}\varepsilon(\hat{b}_{-}^\dagger
e^{-i\omega_m t}+\hat{b}_{-}e^{i\omega_m t}),\label{e10}
\end{eqnarray}
\begin{figure}[t]
\centering
\includegraphics[width=8.4cm]{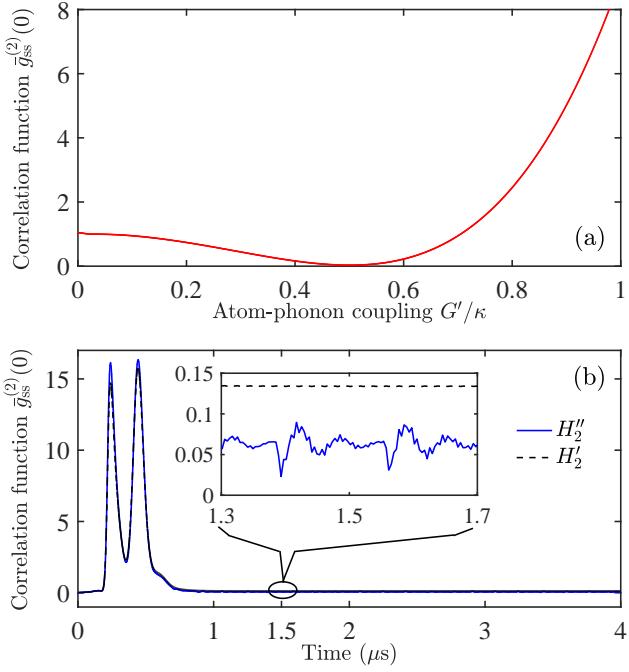}
\caption{(Color online) The second-order correlation function
$\bar{g}_{\rm ss}^{(2)}(0)$, given by Eq.~(\ref{e18}) for the
mechanical supermode $\hat b_-$, versus: (a) the atom-phonon
coupling strength $G'$ and (b) time $t$, in the two-cavity system.
Here $G'=0.5\kappa$, $\sqrt{n_{+}}=51$, $\sqrt{n_{-}}=1$, and
$\varepsilon=0.03\kappa$, while all the other parameters are
the same as in Fig.~\ref{fig:2}.} \label{fig:6}
\end{figure}
where the contribution of
\begin{eqnarray}
\hat{\delta}_b=\frac{g_0}{\sqrt{2}}(\hat{b}^\dagger +\hat{b}),
\quad|\delta_b|\ll J, \label{e11}
\end{eqnarray}
can be neglected safely in the frequencies of optical supermodes.
And \begin{subequations} \label{e12}
\begin{align}
\hat{\alpha}&=J\Big[\Big(\sqrt{{\hat{\delta}_b}^2+J^2}+\hat{\delta}_b\Big)^2+J^2\Big]^{-1/2},\label{e12a}
\\
\hat{\beta}&=\frac{\hat{\alpha}}{J}\Big(\sqrt{{\hat{\delta}_b}^2+J^2}+\hat{\delta}_b\Big).\label{e12b}
\end{align}
\end{subequations}
Meanwhile, we expand $\hat{\alpha}$ and $\hat{\beta}$ to first
order in the small parameter $\hat{\delta}_b/J$ to obtain:
\begin{equation} \label{e13}
\hat{\alpha}=\frac{1}{\sqrt{2}}\Big(1-\frac{\hat{\delta}_b}{2J}\Big),
\quad
\hat{\beta}=\frac{1}{\sqrt{2}}\Big(1+\frac{\hat{\delta}_b}{2J}\Big).
\end{equation}
Inserting Eq.~(\ref{e13}) into Hamiltonian~(\ref{e10}), we obtain
the following tripartite atom-photon-phonon interaction term
\begin{equation}
  \gamma(\hat{b}_{-}^\dagger+\hat{b}_{-})[\hat{\sigma}_{+}(\hat{a}_{+}-\hat{a}_{-})+\hat{\sigma}_{-}(\hat{a}_{+}-\hat{a}_{-})],
 \label{new2}
\end{equation}
with the effective interaction rate $\gamma=gg_0/(4J)$. This
illustrates that we can derive the atom-photon-phonon interaction
from the standard optomechanical coupling.

Let us rewrite the optical supermode operators as
$\hat{a}_{+}=\sqrt{n_{+}}+\delta \hat{a}_{+}$ and
$\hat{a}_{-}=\sqrt{n_{-}}+\delta \hat{a}_{-}$. After inserting
these operators and Eq.~(\ref{e13}) into Hamiltonian~(\ref{e10}),
under the rotating-wave approximation, the Hamiltonian reduces to
\begin{eqnarray}
\hat{H}'_{2}/\hbar&=&-G'(\hat{\sigma}_{+}\hat{b}_{-}+\hat{\sigma}_{-}\hat{b}_{-}^\dagger)+\frac{\varepsilon}{\sqrt{2}}(\hat{b}_{-}^\dagger+\hat{b}_{-})
\nonumber\\&&
+\frac{g}{\sqrt{2}}\frac{G'}{\gamma}(\hat{\sigma}_{+}e^{i\omega_m
t}+\hat{\sigma}_{-}e^{-i\omega_m t}), \nonumber\\&& \label{e14}
\end{eqnarray}
where $G'=\gamma(\sqrt{n_{+}}-\sqrt{n_{-}})$, is the
atom-phonon coupling strength. Note that the atom-phonon
interaction is not only modulated by the average photon numbers of
the supermodes, but also by the atom-photon-phonon interaction
rate $\gamma$. Since $\sqrt{n_{+}}$ is much larger than
$\sqrt{n_{-}}$, we can efficiently enhance the atom-phonon
coupling through the average photon numbers of the supermodes.
Also, the atom-photon-phonon coupling rate here is related to the
optomechanical interaction rate $g_0$ and the cavities-coupling
rate $J$ when the atom-photon interaction rate is fixed. It is
obvious that the atom-photon-phonon interaction can reach a very
large value if $g_0$ is large and $J$ is small enough (for
feasible values of parameters) in the system. The results are
shown in Fig.~\ref{fig:5}: keeping $J$ fixed, the tripartite
interaction rate varies linearly with the optomechanical
interaction and inversely proportional to the cavities coupling
when other parameters are fixed. These illustrate that the
atom-phonon interaction can be controlled with more parameters in
the two-cavity system. Simultaneously, it also improves the
feasibility of future experiments exploring this regime.
\begin{figure*}
\centering
\includegraphics[width=16cm]{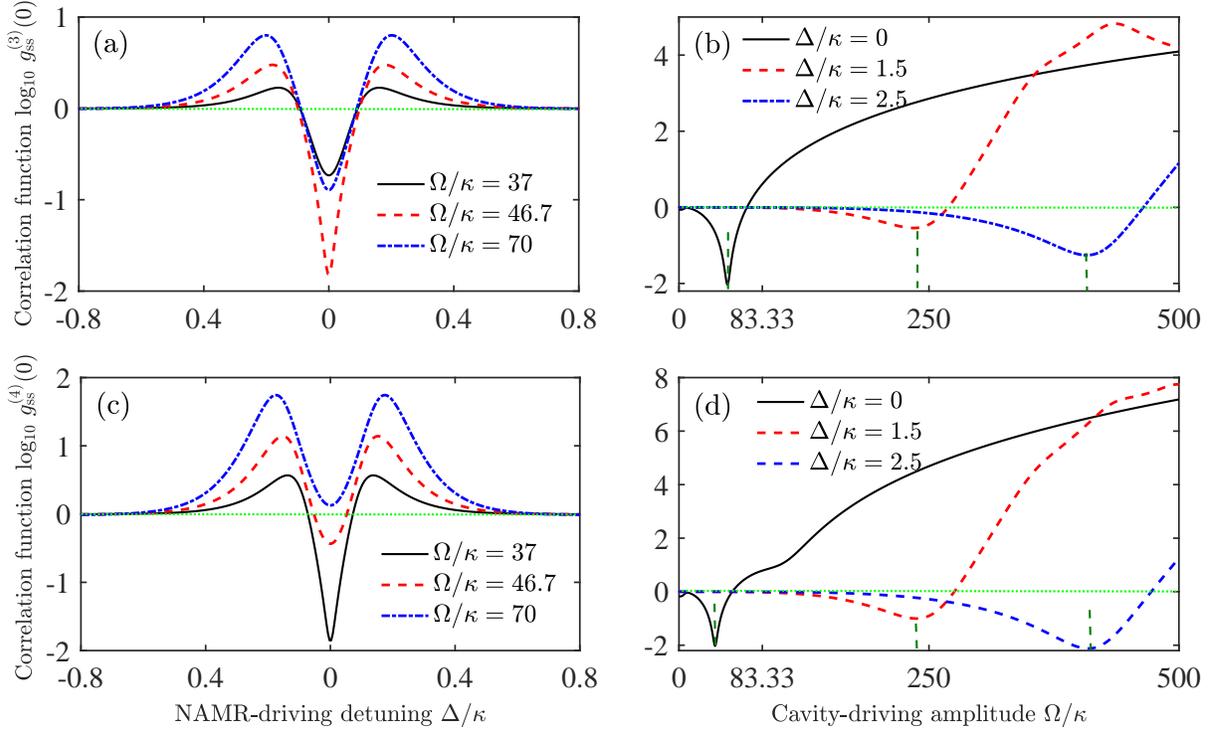}
\caption{(Color online) (a,b) Third- and (c,d) fourth-order
correlation functions $\log_{10}\cor{n}$ of the generated phonon
steady states versus: (a,c) the detuning $\Delta$ between the NAMR
and the corresponding driving field and (b,d) the optical driving
amplitude $\Omega$ under different conditions in the one-cavity
system. All the parameters are the same as in Figs.~\ref{fig:3}
and \ref{fig:4}.} \label{fig:7}
\end{figure*}
\begin{figure}
\centering
\includegraphics[width=8.4cm]{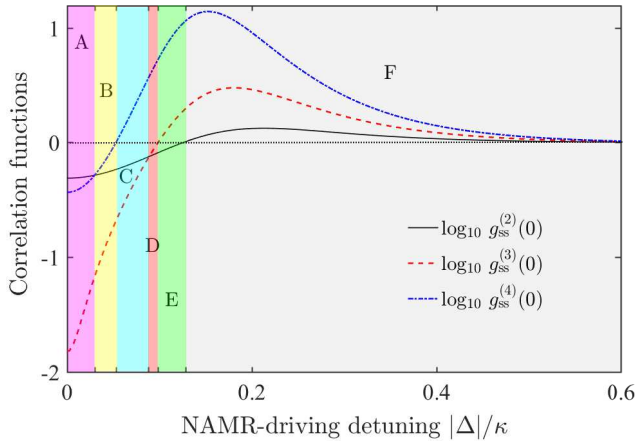}
\caption{(Color online) Correlation functions $\log_{10}\cor{n}$
for $n=2,3,4$ of the generated phonon steady states as a function
of the absolute value of the detuning $|\Delta|$ between the NAMR
and the corresponding driving field in the one-cavity system. We
assume that $\Omega/\kappa=46.7$ and all the other parameters are
the same as in Fig.~\ref{fig:7}. The regions marked $A,...,F$
correspond to different kinds of PB and phonon-induced tunneling,
as summarized in Table~I.} \label{fig:8}
\end{figure}

\begin{table*}
\begin{center}
\caption{Different kinds of phonon blockade (PB) corresponding to the regions of
different detunings shown in Fig.~\ref{fig:8}. Criteria for
non-standard PB and phonon-induced tunneling are given in
Eqs.~(\ref{NPB}) and~(\ref{QT}), respectively.}
\begin{tabular}{c c c c c}
\hline\hline
No. &\,\,\,\,\,\,\,\,\,\, Region &\,\,\,\,\,\,\,\,\,\, Correlation functions &\,\,\,\,\,\,\,\,\,\, Effect &\,\,\,\,\,\,\,\,\,\, Detuning $|\Delta|/\kappa$ \\
\hline
1 &\,\,\,\,\,\,\,\,\,\, A &\,\,\,\,\,\,\,\,\,\, $1>\cor{2}>\cor{4}>\cor{3}$ &\,\,\,\,\,\,\,\,\,\, standard PB     &\,\,\,\,\,\,\,\,\,\, (0, 0.030)      \\
2 &\,\,\,\,\,\,\,\,\,\, B &\,\,\,\,\,\,\,\,\,\, $1>\cor{4}>\cor{2}>\cor{3}$ &\,\,\,\,\,\,\,\,\,\, standard PB     &\,\,\,\,\,\,\,\,\,\, (0.030, 0.053)  \\
3 &\,\,\,\,\,\,\,\,\,\, C &\,\,\,\,\,\,\,\,\,\, $\cor{4}>1>\cor{2}>\cor{3}$ &\,\,\,\,\,\,\,\,\,\, non-standard PB &\,\,\,\,\,\,\,\,\,\, (0.053, 0.088)  \\
4 &\,\,\,\,\,\,\,\,\,\, D &\,\,\,\,\,\,\,\,\,\, $\cor{4}>1>\cor{3}>\cor{2}$ &\,\,\,\,\,\,\,\,\,\, non-standard PB &\,\,\,\,\,\,\,\,\,\, (0.088, 0.098)  \\
5 &\,\,\,\,\,\,\,\,\,\, E &\,\,\,\,\,\,\,\,\,\, $\cor{4}>\cor{3}>1>\cor{2}$ &\,\,\,\,\,\,\,\,\,\, non-standard PB &\,\,\,\,\,\,\,\,\,\, (0.098, 0.124)  \\
6 &\,\,\,\,\,\,\,\,\,\, F &\,\,\,\,\,\,\,\,\,\, $\cor{4}>\cor{3}>\cor{2}>1$ &\,\,\,\,\,\,\,\,\,\, phonon-induced tunneling&\,\,\,\,\,\,\,\,\,\, (0.124, 0.6)    \\
\hline\hline
\end{tabular}
\end{center}
\end{table*}

Under the condition \begin{eqnarray}
\frac{g}{\omega_m}(\sqrt{n_{+}}-\sqrt{n_{-}})\ll1, \label{e16}
\end{eqnarray}
the third term in the Hamiltonian given in Eq.~(\ref{e14})
is quickly oscillating and can be safely omitted, which results in
\begin{eqnarray}
\hat{H}''_{\rm 2}/\hbar
=-G'(\hat{\sigma}_{+}\hat{b}_{-}+\hat{b}_{-}^\dagger
\hat{\sigma}_{-})+\frac{1}{\sqrt{2}}\varepsilon(\hat{b}_{-}^\dagger+\hat{b}_{-}).
\label{e17}
\end{eqnarray}

Just as in the one-cavity system, we demonstrate the property of
the steady-state phonon statistics. The property of phonon
statistics is expressed by a steady-state equal-time second-order
correlation function for the mechanical supermode $\hat{b}_{-}$,
as
\begin{eqnarray}
\label{e18} \bar{g}_{\rm ss}^{(2)}(0)=\frac{\langle
\hat{b}_{-}^{\dagger}\hat{b}_{-}^{\dagger}\hat{b}_{-}\hat{b}_{-}\rangle}{\langle
\hat{b}_{-}^{\dagger}\hat{b}_{-}{\rangle}^2}.
\end{eqnarray}
Here the mechanical supermode $\hat{b}_{-}$ is the linear
superposition of the mechanical operators $\hat{b}_{l}$ and
$\hat{b}_{r}$. As for the two driven and coupled
semipermeable resonators, phonon blockade can be observed even
more easily than that in the one-cavity system. This results from
multipath interference, which is the mechanism of UPB.

We can clearly verify the theory analysis shown above via
numerical calculations. As shown in Fig.~\ref{fig:6}(a), the
strongest phonon blockade appears still at the same optimal point,
which coincides with the corresponding result in the
one-cavity system. This proves that the quantum characteristics
of the mechanical modes and the mechanical supermodes are not
changed in the one- or two-cavity systems. To check the validity
of the approximation from $\hat{H}'_{\rm 2}$ to $\hat{H}''_{\rm
2}$, we show in Fig.~\ref{fig:6}(b) the evolution of the
correlation functions $\bar{g}_{\rm ss}^{(2)}(0)$ governed by the
Hamiltonians $\hat{H}'_{\rm 2}$ and $\hat{H}''_{\rm 2}$. From the
inset of Fig.~\ref{fig:6}(b), we can see that the blue curve
(corresponding to $\hat{H}''_{\rm 2}$) is slightly lower than
exact numerical calculation, the black dashed curve (corresponding
to $\hat{H}'_{\rm 2}$), illustrating that we have made a valid
approximation. Note that these slight differences are visible
only in the magnified inset of Fig.~\ref{fig:6}(b). Moreover, the
steady states of the studied evolution almost coincide with the
value in Fig.~\ref{fig:6}(a) at the optimal point.

In the two-cavity system, Hamiltonian~(\ref{e9}), includes direct
atom-photon and photon-phonon interaction terms. The atom and NAMR
interact with each other indirectly through the common cavity
field (a quantum bus). Thus, we can interpret the predicted phonon
blockade in the two-cavity system as induced by an indirect
atom-phonon interaction. Such an interpretation of phonon blockade
for a single-cavity system, would be less justified, because the
original Hamiltonian in Eq.~(\ref{e1}), already includes a direct
atom-photon-phonon interaction. Note that the effective
Hamiltonian, given in Eq.~(\ref{e17}), also includes apparently a
direct tripartite atom-photon-phonon term. This interaction can be
interpreted as a bipartite atom-phonon interaction, when the
influence of the cavity field is treated as a classical parameter
and incorporated in the coupling constant $G'$.

\section{Higher-order phonon-number correlations}

It is known that not all phonon states with ${g}^{(2)}_{\rm
{ss}}(0)<1$ correspond to standard phonon
blockade~\cite{Hamsen17}. By analyzing higher-order correlation
functions, we can perform a more insightful analysis of PB
generated in our models. Because the physical mechanisms of PB are
very similar in the one- and two-cavity systems, we only analyze
here the one-cavity model.

Higher-order phonon-number steady-state correlation functions can
be defined by:
\begin{eqnarray}
g_{\rm ss}^{(n)}(0)=\frac{\langle
(\hat{b}^{\dagger})^n\hat{b}^n\rangle}{\langle
\hat{b}^{\dagger}\hat{b}{\rangle}^n}, \label{g_n}
\end{eqnarray}
as a generalization of ${g}^{(2)}_{\rm {ss}}(0)$. We recall that
standard PB occurs if ${g}^{(n)}_{\rm {ss}}(0)<1$, not only for
$n=2$ but also for higher orders $n$. Thus, one can talk about
\emph{non-standard} PB if
\begin{equation}
  {g}^{(2)}_{\rm {ss}}(0) < 1\quad {\rm and}\quad {g}^{(n)}_{\rm {ss}}(0)>1
 \label{NPB}
\end{equation}
for some $n>1$. For numerical convenience, we refer to standard PB
if $\cor{n}<1$ for $n=2,3,4$ only. Of course, one can find that in
some cases such PB is not really standard by calculating $\cor{n}$
for $n\ge 5$. Moreover, one can find other classifications of
standard and non-standard PB by applying more refined criteria,
like those in Refs.~\cite{Carreno16, Hamsen17, Miranowicz10}).
Note that PB described by Eq.~(\ref{NPB}) is also referred to as
\emph{unconventional} PB (regime)~\cite{Radulaski17}. It is
clear that this meaning of UPB refers to unconventional properties
of the generated light via PB. Recall that the term UPB used in the
title of this paper and explained in its Introduction refers to an
unconventional mechanism (based on multipath interference) for
creating or enhancing PB~\cite{Gu17,Flayac2017}. Thus, to avoid
confusion, we use the term non-standard PB (rather than UPB) for
mechanical states described by Eq.~(\ref{NPB}).

Moreover, we refer to \emph{phonon-induced tunneling} if the
following conditions are satisfied
\begin{equation}
  \cor{4}>\cor{3}>\cor{2}>1.
 \label{QT}
\end{equation}
This effect is an analogue of \emph{photon}-induced tunneling
studied in, e.g., Ref.~\cite{Rundquist14} (see also references
cited therein). Note that Ref.~\cite{Rundquist14} defines this
photon tunneling via a weaker condition [i.e.,
$\cor{3}>\cor{2}>1$]. If $\cor{4}<1$ [$\cor{5}<1$], one can
interpret such tunneling as three-PB (four-PB), in analogy to
Refs.~\cite{Hamsen17,Miranowicz13}. For simplicity, in this work,
we have neither calculated $\cor{n}$ for $n=5,...$ nor searched
for multi-PB.

Our numerical results for $\cor{3}$ and $\cor{4}$ are shown in
Fig.~\ref{fig:7}, which can be easily compared with the results
for $\cor{2}$ shown in Figs.~\ref{fig:3} and~\ref{fig:4}. A more
explicit comparison of $\cor{n}$ (for $n=2,3,4$) as a function of
the detuning $|\Omega|$ (in units of the atom-damping rate
$\kappa$) is shown in Fig.~\ref{fig:8}. These results, as
summarized in Table~I, reveal regions of specific interrelations
between the correlation functions $\cor{n}$ of different orders
$n$. Each such region corresponds to a specific
phonon-number-correlation effect including standard and
nonstandard PB, as well as phonon-induced tunneling of phonons.

\section{Conclusions}

We have provided a method to generate strong phonon blockade in
hybrid one-cavity and two-cavity optomechanical systems, which are
originally in the weak-coupling regime. Our approach utilizes a
modulated atom-phonon interaction obtained from the tripartite
atom-photon-phonon interaction. We have shown that this
atom-phonon coupling strength can be enhanced by adjusting the
driving-field amplitude and frequency, in the single-cavity
optomechanical system, and the mechanical coupling strength in the
two-cavity case. We presented optimal parameter regimes for
realizing high-quality phonon blockade. Based on the original
Hamiltonian, we also have verified the validity of the
approximations used in our work. By comparing phonon-number
correlation functions $\cor{n}$ of different orders $n=2,3,4$, for
given mechanical steady states in the one-cavity system, we found
different types of standard and non-standard phonon blockade, as
well as phonon-induced tunneling.

This study provides a new route to produce high-quality
phonon blockade using the nonlinearity induced by an atom
(qubit)~\cite{PB16,Singh10,Nonlin13,qubit-resontor10,Cleland04,Pgate17}
in optomechanical hybrid systems, and has potential
applications for quantum technologies.

\begin{acknowledgements}
This work is supported by the National Key R\&D Program of China
grant 2016YFA0301203, the National Science Foundation of China
(Grant Nos. 11374116, 11574104 and 11375067). FN is partially
supported by the MURI Center for Dynamic Magneto-Optics via the
AFOSR Award No.~FA9550-14-1-0040, the Army Research Office (ARO)
under grant number 73315PH, the AOARD grant No.~ FA2386-18-1-4045,
the CREST Grant No.~JPMJCR1676, the IMPACT program of JST, the
RIKEN-AIST Challenge Research Fund, and the JSPS-RFBR grant
No.~17-52-50023. AM and FN acknowledge also a grant from the John
Templeton Foundation.
\end{acknowledgements}

\bibliography{bib_phononblockade}

\end{document}